\documentclass{aa}
\usepackage{graphicx}
\usepackage{natbib}
\usepackage{txfonts}
\bibpunct{(}{)}{;}{a}{}{,} 

\def\cyg{Cygnus~X-3}
\def\msol{$M_{\odot}$}

\begin{document}

\title{Absorption of high-energy gamma rays in Cygnus~X-3}

\titlerunning{Absorption of high-energy gamma rays in Cygnus~X-3}

\author{
B. Cerutti \inst{1,2}
\and G. Dubus \inst{1}
\and J. Malzac \inst{3,4}
\and A. Szostek \inst{5,6}
\and R. Belmont \inst{3,4}
\and A. A. Zdziarski \inst{7}
\and G. Henri \inst{1}
}
\authorrunning{B. Cerutti et al.}

\institute{
UJF-Grenoble 1 / CNRS-INSU, Institut de Plan\'etologie et d'Astrophysique de Grenoble (IPAG) UMR 5274, Grenoble, F-38041, France
\and
Center for Integrated Plasma Studies, Physics Department, University of Colorado, Boulder, CO 80309, USA
\and
Universit\'e de Toulouse; UPS-OMP; IRAP; Toulouse, France
\and
CNRS; IRAP; 9 Av. Colonel Roche, BP 44346, F-31028 Toulouse C\'edex 4, France
\and
Kavli Institute for Particle Astrophysics and Cosmology, SLAC National Accelerator Laboratory, 2575 Sand Hill Road, Menlo Park, CA 94025, USA
\and
Astronomical Observatory, Jagiellonian University, Orla 171, 30-244 Krak\'ow, Poland
\and
Centrum Astronomiczne im. M. Kopernika, Bartycka 18, 00-716 Warszawa, Poland}

\date{Draft \today}
\abstract
{The microquasar Cygnus X-3 was detected at high energies by the gamma-ray space telescopes \emph{AGILE} and \emph{Fermi}. The gamma-ray emission is transient, modulated with the orbital period and seems related to major radio flares, i.e. to the relativistic jet. The GeV gamma-ray flux can be substantially attenuated by internal absorption with the ambient X-rays.}
{In this study, we examine quantitatively the effect of pair production in Cygnus X-3 and put constraints on the location of the gamma-ray source.}
{Cygnus X-3 exhibits complex temporal and spectral patterns in X-rays. During gamma-ray flares, the X-ray emission can be approximated by a bright disk black body component and a non-thermal tail extending in hard X-rays, possibly related to a corona above the disk. We calculate numerically the exact optical depth for gamma rays above a standard accretion disk. Emission and absorption in the corona are also investigated.}
{GeV gamma rays are significantly absorbed by soft X-rays emitted from the inner parts of the accretion disk. The absorption pattern is complex and anisotropic. Isotropization of X-rays due to Thomson scattering in the companion star wind tends to increase the gamma-ray opacity. Gamma rays from the corona suffer from strong absorption by photons from the disk and cannot explain the observed high-energy emission, unless the corona is unrealistically extended.}
{The lack of absorption feature in the GeV emission indicates that high-energy gamma rays should be located at a minimum distance $\sim 10^8$-$10^{10}~$cm from the compact object. The gamma-ray emission is unlikely to have a coronal origin.}

\maketitle

\section{Introduction}

Gamma-ray emission from \cyg\ was detected by the gamma-ray space telescopes {\em AGILE} and {\em Fermi} \citep{2009Natur.462..620T,2009Sci...326.1512F}. This system is formed of a Wolf-Rayet star \citep{1996A&A...314..521V} and an unidentified compact object (a black-hole or a neutron star) in a very tight 4.8~h orbit, situated at 7-9~kpc \citep{1983ApJ...273L..71D,2000A&A...357L..25P,2009ApJ...695.1111L}. \cyg\ is also known as a microquasar as there is clear evidence of resolved relativistic radio jets \citep{2000ApJ...545..939M,2001ApJ...553..766M}. The gamma-ray emission is transient, modulated with the orbital period of the system \citep{2009Sci...326.1512F} and seems to be related to giant radio flares and episodes of major ejections of a relativistic jet. The spectrum measured by {\em Fermi} is a soft power-law of index $\sim 2.7$ with no sign of high energy cut-off, for a total luminosity above 100~MeV of about $3\times 10^{36}~$erg~s$^{-1}$ (at 7~kpc). The gamma-ray orbital modulation was interpreted as the result of Compton upscattering of thermal photons from the Wolf-Rayet star by relativistic pairs accelerated in the jet \citep{2009Sci...326.1512F,2010MNRAS.404L..55D}.

The propagation of energetic gamma rays to the observer can be greatly affected by pair production $\gamma+\gamma\rightarrow e^{+}+e^{-}$ with the ambient radiation field in the system, regardless the underlying production mechanisms for these gamma rays. This issue motivated many different studies in the past, at a time when \cyg\ was thought to emit GeV, TeV and even PeV gamma-ray photons (see for instance \citealt{1983ApJ...271..304V,1983ApJ...274L..23G,1984ApJ...287..338A,1984A&A...133...80C,1987ApJ...322..838P,1992A&A...264..127C,1993Ap&SS.204..141W}). \citet{1987A&A...175..141H} invoked the effect of gamma-ray absorption as a possible explanation for the non-detection of Cygnus X-3 at GeV energies by {\em COS-B}. This question is quantitatively revisited here in the context of the new gamma-ray observations of \cyg. The goal of this article is to formulate constraints on the location of the gamma-ray source in the system.

High-energy gamma rays in the energy band $\epsilon_1=100~$MeV-$10~$GeV annihilate with soft X-rays photons of energy $\epsilon_{\rm{min}}\sim m_e^2 c^4/\epsilon_1=25~$eV~-~$2.5~$keV. Hence, the effect of the gamma-ray opacity at high energy depends on the ambient X-ray density in the system. Known Wolf-Rayet stars have effective temperatures around $T_{\star}\sim 10^5~$K (see {\em e.g.} the review by \citealt{2007ARA&A..45..177C}) and provide a high density of $\sim 20~$eV $<\epsilon_{\rm{min}}$ thermal radiation (with a bolometric luminosity $L_{\star}\gtrsim 10^{38}~$erg~s$^{-1}$). Absorption with stellar radiation is then irrelevant in this context. Note that the star would have a major role in the optical depth for very-high energy ($>100~$GeV) gamma rays \citep{1984MNRAS.211..559F,1987ApJ...322..838P,1993MNRAS.260..681M,2010MNRAS.406..689B}. Some Wolf-Rayet stars present an excess in soft X-rays with luminosities up to $\sim 10^{33}$ erg~s$^{-1}$, particularly in nitrogen-rich stars ({\em e.g.} \citealt{2010AJ....139..825S}). This source of photons would be negligible in \cyg\ where the X-ray luminosity is about $10^{38}$ erg~s$^{-1}$, and will not be considered here. If the accreting compact object is a neutron star, thermal radiation is emitted by the surface of the star. With a typical surface temperature $T_{\rm{ns}}\sim 10^6~$K (see {\em e.g.} \citealt{2004ARA&A..42..169Y}), the neutron star provides $\sim 0.25~$keV soft X-rays able to absorb 1~GeV photons. The luminosity of the neutron star $L_{\rm{ns}}\approx 10^{33}~$erg~s$^{-1}$ is too small and will be neglected as well in the following. The dominant source of soft X-rays in \cyg\ is provided by the accretion disk formed around the compact object. The inner parts of the disk could be very hot with temperature $T_{\rm{disk}}\sim 10^6$-$10^7$~K and emit copious $\sim~1~$keV soft X-rays in the system. We provide in Sect.~\ref{sect_disk} and in Appendix~\ref{app_a} a complete study of the absorption of GeV photons in the radiation field of an accretion disk.

Long-term observations in X-rays revealed that \cyg\ presents complex spectral and temporal features. \citet{2008MNRAS.388.1001S} classified the X-ray spectral states into five distinct groups, from the `hard state' to the `soft state'. In the hard state, the X-ray flux is dominated by hard X-rays ($>10~$keV) and the spectrum is well-fitted by comptonized emission on thermal and non-thermal electrons with a cut-off at $\approx 20~$keV and a Compton reflection component. The soft state is dominated by the bright thermal emission from the accretion disk in soft X-rays ($\sim 1~$keV), and contains also a non-thermal tail in hard X-rays extending at least up to 100~keV. Gamma-ray flares were observed only when \cyg\ was in the soft state (during major radio flares). Therefore, we focus only on the soft state and examine whether the non-thermal X-ray component could, in addition to the accretion disk, contribute significantly to increase the gamma-ray opacity in \cyg\ (Sect.~\ref{sect_corona}). The nature and the origin of this non-thermal emission is unknown. This is possibly related to electrons in the hot corona above the disk or in the relativistic jet. The extrapolation of the hard X-ray tail in the soft state up to the GeV energy domain is tempting. In Sect.~\ref{sect_corona}, we examine also the possibility that the gamma-ray emission comes from a hot corona above the accretion disk using the one-zone radiative model {\tt BELM} described in \citet{2008A&A...491..617B}. Gamma-ray emission from a magnetized corona was also investigated by \citet{2010A&A...519A.109R} with an application to the microquasar Cygnus~X-1.

\begin{figure}
\centering
\resizebox{\hsize}{!}{\includegraphics{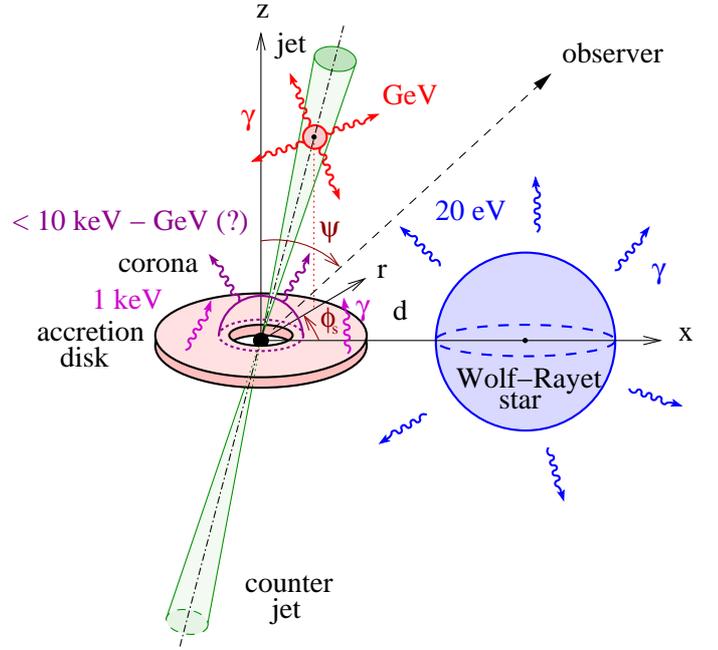}} 
  \caption{Diagram of the microquasar \cyg\ with the different sources of photons considered in this study. The observer sees the disk at an angle $\psi$. In cylindrical coordinates, the gamma-ray source is located at $(r,\phi_{\rm{s}},z)$, possibly in the relativistic jet. For a close-up view of the disk see the appendix, Fig.~\ref{fig_disk}.}
\label{microquasar}
\end{figure}

The global and simplified picture of \cyg\ adopted in this paper is summarized and depicted in Fig.~\ref{microquasar}. The main results and conclusions of this study are exposed in the last section (Sect.~\ref{sect_ccl}).

\begin{figure}
\centering
\resizebox{\hsize}{!}{\includegraphics{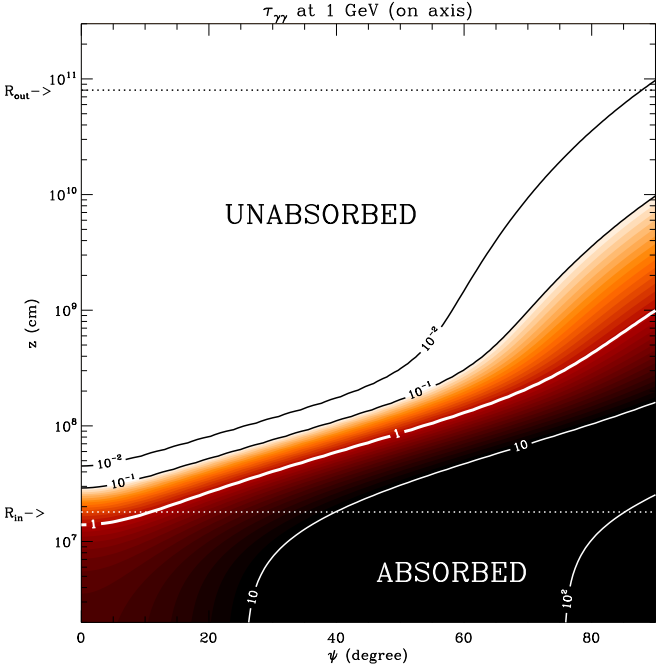}} 
  \caption{Optical depth $\tau_{\gamma\gamma}$ map for a gamma-ray photon of energy $\epsilon_1=1~$GeV, integrated along the line of sight up to the observer situated at infinity. The compact object is a $M_{\rm{co}}=20~$\msol\ black hole accreting at $\dot{M}_{\rm{acc}}=10^{-8}~$\msol~yr$^{-1}$. The inner radius of the disk is $R_{\rm{in}}=6~r_g\approx 1.8\times 10^7~$cm and the outer radius is $R_{\rm{out}}=8\times 10^8~$cm. The altitudes $z\equiv R_{\rm{in}}$ and $z\equiv R_{\rm{out}}$ are indicated by dotted lines. Absorption is calculated with 
  the temperature profile given in Eq.~(\ref{temp_profile_gr}). 
  Primary gamma rays are injected along the axis of the disk at an altitude $z$ ($r\equiv 0$). The observer sees the disk with an inclination $\psi$. Bright regions indicate low opacity or `UNABSORBED' regions, and dark regions high opacity or `ABSORBED' regions. Level lines $\tau_{\gamma\gamma}=10^{-2},~10^{-1},~1,~10,~10^2$ are overplotted.}
\label{fig_anis}
\end{figure}

\section{Absorption by the accretion disk}{\label{sect_disk}}

In this section, we propose a full study of gamma-ray absorption by the photons from an accretion disk formed around the compact object (Fig.~\ref{microquasar}). With simple assumptions on the physics and geometry of the disk (\S\ref{sect_sa}), we derive the main properties of the optical depth for gamma rays (\S\ref{sect_anis}, \ref{sect_map}) with an application to \cyg. Soft X-rays emitted by the disk can be significantly scattered and absorbed by the dense Wolf-Rayet star wind in \cyg. We discuss in the last section (\S\ref{sect_wind}) the impact of the stellar wind on the gamma-ray opacity.

\subsection{Accretion disk model}\label{sect_sa}

Pair production of high-energy gamma rays in the radiation field of an accretion disk has been investigated by several authors in the context of X-ray binaries \citep{1992A&A...264..127C,1993A&A...278..307B}, but also in AGN (see {\em e.g.} \citealt{1995ApJ...453...83B,1997ApJ...475..534Z,2008MNRAS.391..624S}). In most of these previous studies, the accretion disk is optically thick, geometrically thin, flat and follows the standard quasi-Keplerian disk-$\alpha$ model of \citet{1973A&A....24..337S}. According to this theory, each surface element of the disk is at thermal equilibrium. In the non-relativistic case and if the boundary conditions of the disk are neglected, the radial profile of the effective temperature follows
\begin{equation}
T\left(\bar{R}\right)=T_{\rm{g}}\bar{R}^{-3/4},
\label{temp_profile}
\end{equation}
where $\bar{R}=R/r_{\rm{g}}$. $R$ is the distance from the center of the disk and $r_{\rm{g}}=G M_{\rm{co}}/c^2$ is the gravitational radius of the compact object. $T_{\rm{g}}$ is formally the temperature of the disk at $R\equiv r_{\rm{g}}$ which depends on the accretion rate $\dot{M}_{\rm{acc}}$ onto the central compact object of mass $M_{\rm{co}}$; $T_{\rm g}$ is defined as
\begin{equation}
T_{\rm{g}}=\left(\frac{3\dot{M}_{\rm{acc}}c^6}{8\pi\sigma_{\rm{SB}}G^2 M_{\rm{co}}^2}\right)^{1/4},
\end{equation}
where $G$ is the gravitational constant and $\sigma_{\rm{SB}}$ is the Stefan-Boltzmann constant. Eq.~(\ref{temp_profile}) can be rewritten as
\begin{equation}
T(R)=7.7\times 10^6~\dot{M}_{8}^{1/4}M_{20}^{1/4}R_{7}^{-3/4}~\rm{K},
\label{temp_profile2}
\end{equation}
where $\dot{M}_{8}=M_{\rm{acc}}/10^{-8}~$\msol~yr$^{-1}$, $M_{20}=M_{\rm{co}}/20~$\msol~and $R_7=R/10^7~$cm. The inner radius of the accretion disk $R_{\rm{in}}$ is not well constrained but is usually set at the last stable orbit ({\em i.e.} $R_{\rm{in}}\equiv 6~r_g$ for a non-rotating black hole) in the soft-state where the accretion disk component is very bright ({\em e.g.} \citealt{1997ApJ...489..865E}). For an accreting neutron star, we will assume that the inner part of the disk reaches the neutron star surface $R_{\rm{in}}\equiv R_{\rm{ns}}= 10~$km. The emission produced by the dissipation of the kinetic energy of the accreted material in the neutron star surface is not considered here. We will do the approximation that the observed X-ray luminosity comes from the disk only.

The temperature profile can be substantially modified close to the compact object by the inner boundary condition of the disk, neglected in Eq.~(\ref{temp_profile}). Using a torque-free inner boundary condition and a pseudo-Newtonian gravitational potential \citep{1980A&A....88...23P}, Eq.~(\ref{temp_profile}) is changed into \citep{1999MNRAS.309..496G}
\begin{equation}
T\left(\bar{R}\right)=T_{\rm{g}}\left[\frac{\bar{R}-2/3}{\bar{R}\left(\bar{R}-2\right)^{3}}\left(1-\frac{3^{3/2}}{2^{1/2}}\frac{\bar{R}-2}{\bar{R}^{3/2}}\right)\right]^{1/4}.
\label{temp_profile_gr}
\end{equation}
For $\bar{R}\lesssim 10$, the temperature becomes significantly smaller than in Eq.~(\ref{temp_profile}) (about a factor 2 at $\bar{R}=10$). Far from the last stable orbit $\bar{R}\gg 6$, this solution matches the profile $T\propto \bar{R}^{-3/4}$. Since the effect of the gamma-ray absorption depends critically on the temperature in the inner parts of the disk, we compute here the optical depth for high-energy gamma rays in the framework of a $\alpha$ type disk with the modified profile temperature given by Eq.~(\ref{temp_profile_gr}). General relativistic effects such as Doppler beaming and the bending of light trajectories are neglected.

In close binaries, the accretion disk does not extend beyond a fraction of the Roche lobe radius due to tidal torques exerted by the companion star, \citep{1977ApJ...216..822P,1977MNRAS.181..441P}. We set the outer radius of the disk $R_{\rm{out}}$ at the Roche lobe radius. The energy of the photons emitted in the outer part of the disk falls in the infra-red band, hence is too small to absorb $100~$MeV - $10~$GeV gamma rays, but could annihilate with TeV photons.

The total luminosity of the disk $L_{\rm{disk}}$ depends on the accretion rate of matter $\dot{M}_{\rm{acc}}$ onto the compact object, such that (for a disk truncated at $R_{\rm{in}}\leq R$ extending at infinity and following the temperature profile in Eq.~\ref{temp_profile}, \citealt{1994hea..book.....L})
\begin{equation}
L_{\rm{disk}}=\frac{G\dot{M}_{\rm{acc}}M_{\rm{co}}}{2R_{\rm{in}}}\approx 8.4\times 10^{37}~\dot{M}_8 M_{20} R_{7}^{-1}~\rm{erg}~\rm{s}^{-1}.
\label{l_disk}
\end{equation}
We assume for simplicity that the total bolometric luminosity observed in X-rays (corrected from absorption by the Wolf-Rayet star wind and the interstellar medium) originates from the disk in \cyg, so that $L_{\rm{X}}\approx L_{\rm{disk}}$. Pairs produced by the absorption of gamma rays could interact with photons from the disk or from the star and develop an electromagnetic cascade, reducing the opacity in the system (see \citealt{1985Ap&SS.115...31A,1992A&A...264..127C,1993Ap&SS.204..141W,2010MNRAS.406..689B}). The cascade might be quenched if the ambient magnetic field is large enough in the system (see Sect.~\ref{sect_corona_gev}). Synchrotron emission from secondary pairs may be important or even dominant in hard X-rays and soft gamma rays \citep{2008A&A...489L..21B,2010A&A...519A..81C,2010A&A...519A.109R}. The cascade above the accretion disk is not considered in the following.

\subsection{Anisotropic effects}\label{sect_anis}

\begin{figure}
\centering
\resizebox{\hsize}{!}{\includegraphics{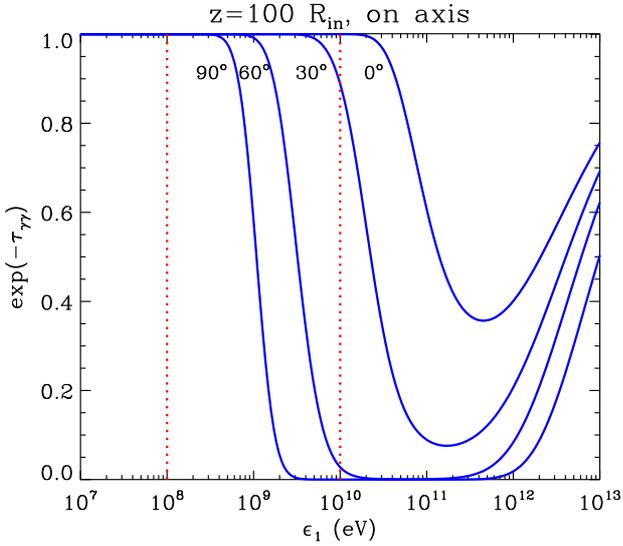}} 
  \caption{Transmitted flux $\exp(-\tau_{\gamma\gamma})$ as a function of the gamma-ray photon energy $\epsilon_1$, for the following viewing angle $\psi=0\degr$ ({\em right}),$~30\degr,~60\degr$ and $90\degr$ ({\em left}). The disk has the same physical properties as in Fig.~\ref{fig_anis}. Gamma-rays are injected at $z=100~R_{\rm{in}}$ for illustrative purpose only. The {\em Fermi}-LAT energy band (100~MeV-10~GeV) is delimited by the dotted lines.}
\label{fig_anis_eps1}
\end{figure}

\begin{figure*}
\centering
\resizebox{18cm}{!}
{\includegraphics*{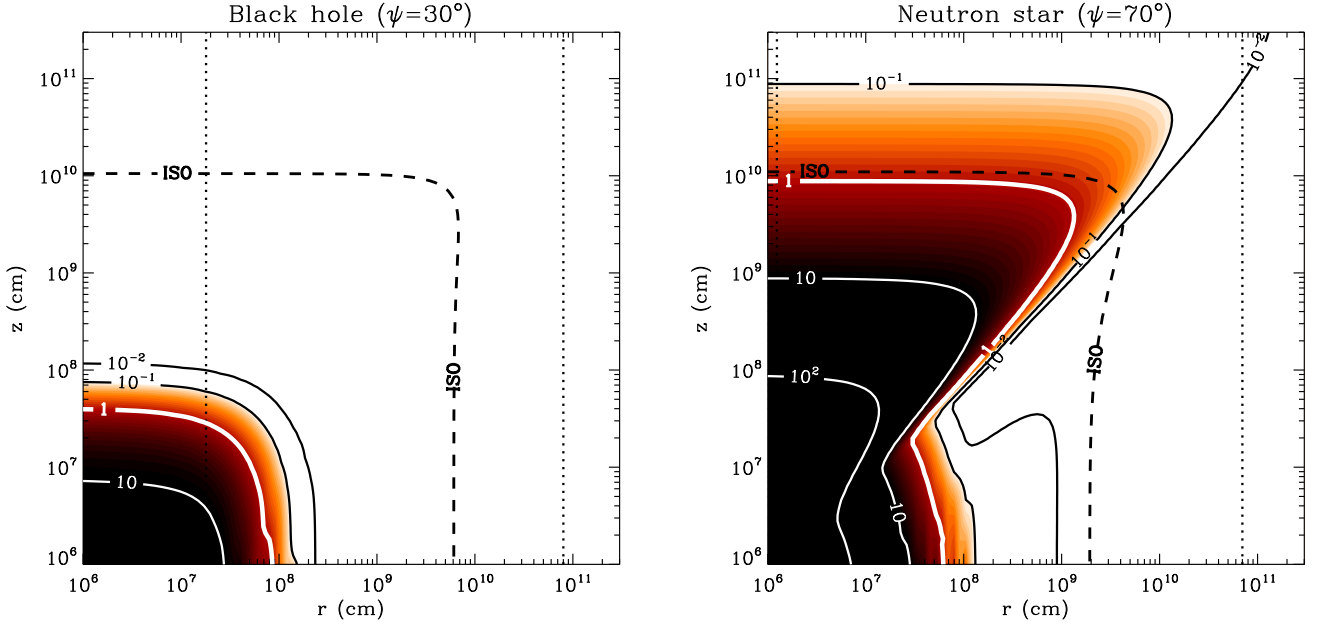}}
  \caption{Optical depth map for 1~GeV photons in \cyg, as a function of the location of the gamma-ray source above the accretion disk of coordinates $(r,\phi_{\rm{s}},z)$. The level lines corresponding to $\tau_{\gamma\gamma}=10^{-2},~10^{-1},~1,~10,~10^2$ are also overplotted. The accreting compact object is a 20~\msol~black hole ({\em i.e.} $\psi=30\degr$, {\em left} panel) or a neutron star ($\psi=70\degr$, {\em right} panel), located at the center of the disk (origin). The dashed line labelled `ISO' shows the contour $\tau_{\gamma\gamma}=1$ in the case where the radiation from the accretion disk is fully isotropized. The maps shown here are calculated for $\phi_{\rm{s}}=0\degr$ (see Fig.~\ref{microquasar}). The inner radius of the disk is $R_{\rm{in}}=6~r_{\rm{g}}$ and the outer radius chosen at the Roche lobe radius for the compact object. Dotted lines indicate the inner and outer radius of the disk.}
\label{fig_map}
\end{figure*}

The radiation field from the accretion disk seen by a gamma ray is highly anisotropic. The cross section for pair production depends on the scattering angle between the gamma ray and the target photon. The gamma ray is injected above the accretion disk at a location given by $(r,\phi_{\rm{s}},z)$ in cylindrical coordinates (see Fig.~\ref{fig_disk}). We compute the total gamma-ray optical depth exactly, {\em i.e.} by including all the geometrical effects. Appendix~\ref{app_a} provides the full details for the calculation of the gamma-ray optical depth $\tau_{\gamma\gamma}$. Fig.~\ref{fig_anis} is the result of the numerical integrations in Eq.~(\ref{tau_gg}), and gives the total optical depth above the accretion disk as a function of the viewing angle $\psi$ and the altitude $z$ where gamma rays are injected. The temperature profile of the disk follows Eq.~(\ref{temp_profile_gr}). In this calculation, the gamma-ray source is on the axis of the disk ($r\equiv 0$, see Figs.~\ref{microquasar}, \ref{fig_disk}) and emits 1~GeV photons. For illustrative purpose, we consider the case of a 20~\msol\ black hole accreting at a rate $\dot{M}_{\rm{acc}}=10^{-8}~$\msol~yr$^{-1}$ (corresponding to $L_{\rm{disk}}\approx 10^{38}$~erg~s$^{-1}$).

The altitude at which the medium becomes transparent to gamma rays is smallest at low inclinations ($\psi\lesssim 30\degr$). The direction of the gamma rays and the photons from the disk are about the same for these inclinations. For higher inclination ($\psi\gtrsim 30\degr$), the GeV photon crosses a thicker radiation field and the collision with the photons from the disk becomes closer to head-on, resulting in a higher opacity for a given altitude $z$. In this case, gamma rays should be injected at $z\gtrsim10^8$-$10^9$~cm for $\psi=90\degr$.

The gamma-ray optical depth depends also highly on the energy of the gamma-ray photon. The map shown in Fig.~\ref{fig_anis} changes significantly with energy, although the same anisotropic pattern remains. Fig.~\ref{fig_anis_eps1} shows this dependence for various inclinations. It can be noticed here that in addition to the massive star, the accretion disk (the outer parts emitting in optical-infrared) can also be a strong source of gamma-ray absorption for TeV gamma-rays (if any, see the upper-limit and the discussion in \citealt{2010ApJ...721..843A}).

\subsection{Map of the gamma-ray optical depth in Cygnus X-3}\label{sect_map}

\begin{figure*}
\centering
\resizebox{18cm}{!}
{\includegraphics*{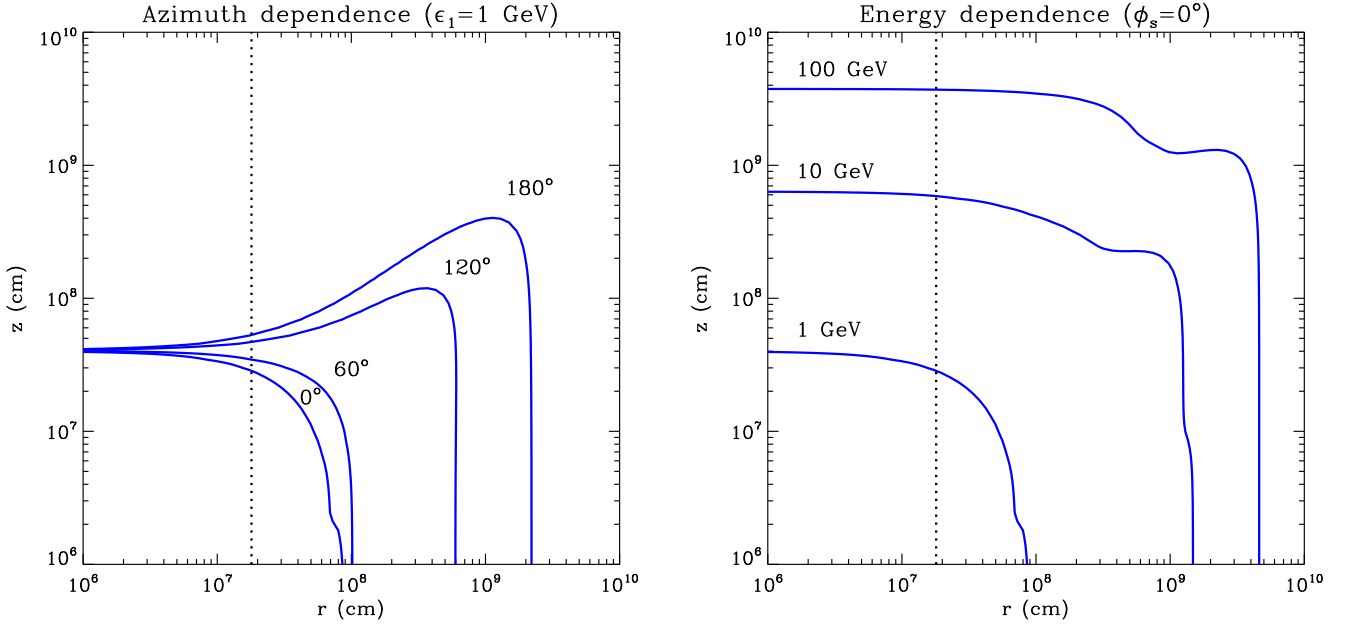}}
  \caption{Azimuth ({\em left} panel) and energy dependences ({\em right} panel) of the gamma-ray photosphere contour ($\tau_{\gamma\gamma}\equiv 1$) above the accretion disk, for a 20~\msol\ black hole ($\psi=30\degr$). In the {\em left} panel, $\phi_{\rm{s}}=0\degr,~60\degr,~120\degr,~$and $180\degr$ with $\epsilon_1=1~$GeV. In the {\em right} panel, the energy of the gamma-ray is $\epsilon_1=1,~10$ and 100~GeV for $\phi_{\rm{s}}=0\degr$. Pair production with stellar radiation is not included in this plot. The dotted line shows the inner radius of the disk.}
\label{fig_map2}
\end{figure*}

The inclination of the system with respect to the observer is unknown in \cyg. We consider the two extreme solutions chosen by \citet{2008MNRAS.386..593S}, based on their estimate of the minimum and maximum mass loss rate of the Wolf-Rayet star. In the first solution the inclination of the system is $i=70\degr$, corresponding to the case where the compact object is a neutron star ($M_{\rm{co}}=1.4~$\msol) orbiting a 5~\msol~Wolf-Rayet star. At the other end, the inclination is $i=30\degr$ and corresponds to a 20~\msol~black hole and a 50~\msol~Wolf-Rayet star companion. These two cases represent also two extreme solutions in terms of gamma-ray absorption as the optical depth is higher for high inclinations (Fig.~\ref{fig_anis}). In addition, these two cases correspond respectively to the maximum and minimum value for the inner radius of the accretion disk, hence for a hotter and cooler accretion disk. Assuming that the disk is in the orbital plane (so that $i\equiv\psi$), we investigate the spatial distribution of the optical depth for GeV photons above the accretion disk.

Fig.~\ref{fig_map} presents the result of the calculation, for the particular case $\phi_s=0\degr$ (Figs.~\ref{microquasar},~\ref{fig_disk}). We take the bolometric X-ray luminosity of \cyg\ as the total luminosity of the accretion disk $L_{\rm{disk}}\approx L_{\rm{X}}$, with $L_{\rm{X}}\approx 10^{38}~$erg~s$^{-1}$ (see {\em e.g.} \citealt{2009A&A...501..679V}) which is about the Eddington luminosity $L_{\rm{Edd}}$ for the neutron star (accretion rate of $\dot{M}_{\rm{acc}}=10^{-8}$~\msol~yr$^{-1}$). In these maps, the line $\tau_{\gamma\gamma}=1$ delimits the contour of the `gamma-ray photosphere' above the disk, {\em i.e} the surface beyond which gamma rays can escape from the system and reach the observer. Note that similar maps were found in the context of AGN \citep{2008MNRAS.391..624S,2010MNRAS.401.1983S}. Absorption is more marked in the neutron case since the inner radius of the disk is $M_{\rm{bh}}/M_{\rm{ns}}\approx 14$ times smaller than in the black hole solution. Fig.~\ref{fig_map2} ({\em left} panel) shows the variations of the gamma-ray photosphere location with the azimuth angle $\phi_{\rm{s}}$. Gamma-ray absorption is stronger for $\phi_{\rm{s}}=180\degr$ than for $\phi_{\rm{s}}=0\degr$ because the gamma ray photon crosses a thicker radiation field from the disk before it reaches the observer. The gamma-ray photosphere is located at a distance between $\sim 10^{10}$-$10^{11}~$cm from the compact object for the neutron star and $\sim 5\times 10^{7}$-$10^{9}~$cm for the black hole at 1 GeV. Because GeV gamma rays seem not absorbed in \cyg, the gamma-ray source should lie beyond this critical distance. The gamma-ray photosphere contour depends also on the energy of the gamma-ray photon (Fig.~\ref{fig_map2}, {\em right} panel).

The transparency constraint depends sensitively on the value of the inner radius of the thin accretion disk, since GeV gamma rays are mostly absorbed by photons emitted in the innermost, hottest region. For thin disks with higher $R_{\rm{in}}$, gamma-ray absorption is lower and is negligible at 1 GeV if
\begin{equation}
R_{\rm{in}}\gtrsim 1.4\times10^8~\dot{M}_{8}^{1/3} M_{20}^{1/3}~\rm{cm},
\end{equation}
{\em i.e.} for disks with such inner radii, the mean energy of produced photons equals the minimal energy required for pair production $2.7~k T\equiv 0.25~$keV (see Eq.~\ref{temp_profile2} for $R_{\rm{in}}\gg 6~r_{\rm{g}}$ and Eq.~\ref{threshold}). The (unspecified) innermost accretion flow might still contribute to gamma-ray attenuation. However, the high X-ray luminosity would be difficult to account for if the thin disk is truncated above $\ga~$50~$r_{\rm{g}}$. The accretion rate would have to be as high as about half the Eddington accretion rate limit for a 20~\msol\ black hole ($\dot{M}_{\rm{Edd}}=L_{\rm{Edd}}/c^2$).

\subsection{X-ray scattering and absorption in the stellar wind}\label{sect_wind}

Thomson scattering and absorption of soft X-rays ($\lesssim 10~$keV) in the dense Wolf-Rayet star wind is a major issue in \cyg. This effect is partly responsible for our poor knowledge of the physical properties of the disk such as the temperature and the instrinsic luminosity, in spite of many years of observations in X-rays \citep{2008MNRAS.386..593S,2008MNRAS.384..278H}. The stellar wind could significantly deplete and redistribute the soft X-ray density seen by the GeV gamma ray, relaxing the constraints formulated in the previous section (Sect.~\ref{sect_map}) on the minimum distance to the accreting object.

We calculate the conservative photoelectric optical depth to estimate the absorption of soft X-rays in the stellar wind. For a neutral, isotropic and radial stellar wind assumed composed of pure helium, the density of helium atoms at a distance $\Lambda$ from the center of the star is
\begin{equation}
n_{\rm{He}}\left(\Lambda\right)=\frac{\dot{M}_{\rm{w}}}{4\pi m_{\rm{H}} \mu_{\rm{i}} \Lambda^2 v_{\rm{w}}}.
\end{equation}
$\dot{M}_{\rm{w}}$ is the mass loss rate of the star, $m_{\rm{H}}$ is the hydrogen atom mass, $v_{\rm{w}}$ the stellar wind velocity and $\mu_{\rm{i}}$ is the mean molecular weight equals to 4 for an helium wind. With a typical mass loss rate $\dot{M}_{\rm{w}}=10^{-5}~$\msol~yr$^{-1}$ and a wind velocity $v_{\rm{w}}=1500~$km~s$^{-1}$ in Wolf-Rayet stars, $n_{\rm{He}}\approx5\times10^{11}~$cm$^{-3}$ at the compact object location. The column density of target atoms $N_{\rm{He}}$ crossed by X-rays from the compact object up to a distance $\rho$ is
\begin{equation}
N_{He}\left(\rho\right)=\int_0^{\rho}n_{He}\left(\Lambda\right)d\rho',
\label{Nhe}
\end{equation}
where $\rho'$ is the path length integration variable between the source of X-rays ($\rho'\equiv 0$) and the distance $\rho$. The distance $\Lambda$ is related to $\rho'$, the orbital separation $d$ and the angle $\theta$ between the X-ray photon direction and the orbital plane (see Fig.~\ref{fig_disk}) {\em via} the relation $\Lambda^2=\rho'^2+d^2-2\rho'd\sin\theta$. At 1~keV, the photoelectric absorption cross section is about $\sigma_{\rm{PE}}(1~\rm{keV})\approx 2\times 10^{-22}~$cm$^2$ \citep{1983ApJ...270..119M} (assuming that the abundance of metals per unit mass in the wind is similar to cosmic abundances). The photoelectric optical depth in the wind is $\tau_{\rm{w}}(\rho)=N_{\rm{H}}(\rho)\sigma_{\rm{PE}}(1~\rm{keV})\gtrsim 1$ if $\rho\gtrsim 10^{10}~$cm (for the black hole and the neutron star case). This critical distance is much bigger than the gamma-ray photosphere radius in the black hole case but is comparable for the neutron star solution. This is probably an overestimate in \cyg\ as the wind should be fully ionized around the X-ray source over distances comparable to the orbital separation $\sim 10^{11}~$cm \citep{2008MNRAS.386..593S}, hence much larger than the extension of the gamma-ray photosphere. We conclude that absorption of soft X-rays in the wind should not affect the gamma-ray opacity above the accretion disk.

Thomson scattering between X-rays and unbound electrons in the wind could result in a redistribution and isotropization of the directions of the photons. For a fully ionized wind, the Thomson optical depth in \cyg\ gives $\tau_T\sim 2 N_{\rm{He}}\sigma_{\rm{T}} \approx$ 0.5-1.8, for $\psi=30\degr$ and $70\degr$ respectively (where the factor 2 accounts for 2 electrons per He atoms, $\sigma_{\rm{T}}$ is the Thomson cross section, and $N_{\rm{He}}$ is the helium column density integrated up to infinity in Eq.~\ref{Nhe}). Gamma rays would see a different angular distribution (more isotropic) of the incoming X-ray distribution. This might increase the gamma-ray optical depth (as collisions between photons would be on average closer to head-on) and reduce the anisotropic pattern of the gamma-ray photosphere in Fig.~\ref{fig_map}, in particular in the neutron star case. To address this question more quantitatively, we performed the same calculation of the gamma-ray optical depth as in the previous section assuming that the angular distribution of the photons from each unit surface of the disk is isotropic (the integrand in Eq.~\ref{tau_gg} is averaged over a uniform distribution of pitch angle $\theta_0$). This extreme situation of full isotropization yields an upper-limit to the effect of X-ray scattering in the stellar wind on the gamma-ray optical depth. Calculations show that the gamma-ray photosphere is nearly spherical of radius $\approx 10^{10}~$cm (Fig.~\ref{fig_map}, dashed lines). The photosphere is about a hundred times more extended in the black hole case. The inclination of the system has a weak impact on this result. The real gamma-ray photosphere would likely be contained between the anisotropic and the isotropic case {\em i.e.} $10^{8}$-$10^{10}$~cm, regardless the nature of the compact object in Cygnus~X-3.

\section{Absorption and emission in the corona}{\label{sect_corona}}

The goal of this part is to investigate the role of a corona in emission and absorption of high-energy gamma rays. First, we estimate the contribution of non-thermal hard X-rays from the corona to the gamma-ray opacity (\S \ref{sect_hard}). Second, we model the emission of the corona and look whether GeV gamma rays can be produced and escape the system in \cyg\ (\S \ref{sect_corona_gev}).

\subsection{Absorption with non-thermal hard X-rays}{\label{sect_hard}}

We modeled the corona as a spherical region of radius $R_{\rm{c}}$ centered at the location of the compact object where energetic electrons are injected and radiate {\em via} synchrotron and inverse Compton scattering. For an optically thin and homogeneous spherical source, the emerging photon density has a radial dependence \citep{1979A&A....76..306G,1985ApJ...298..128B}. We approximate the emission inside the corona ($\rho<R_{\rm{c}}$, where $\rho$ is the distance to the center) as homogeneous and isotropic. Outside ($\rho>R_{\rm{c}}$), photons are assumed to escape radially, hence the flux decreases as $1/\rho^2$ and the corona appears as a point-like emitter. X-ray photons are injected with a power-law energy distribution of index $-2$ ranging from 10~keV to 100~keV, corresponding to a total power injected in the corona of $L_{\rm{c}}=5\times 10^{36}~$erg~$\rm{s}^{-1}$. This simple spectral parametrization is a rather good representation for the hard X-ray tail observed in the soft state above 10~keV (and radio flares, see {\em e.g} \citealt{2008MNRAS.388.1001S} Fig.~8). Appendix~\ref{app_b} gives more details on the model adopted.

\begin{figure}
\centering
\resizebox{\hsize}{!}{\includegraphics{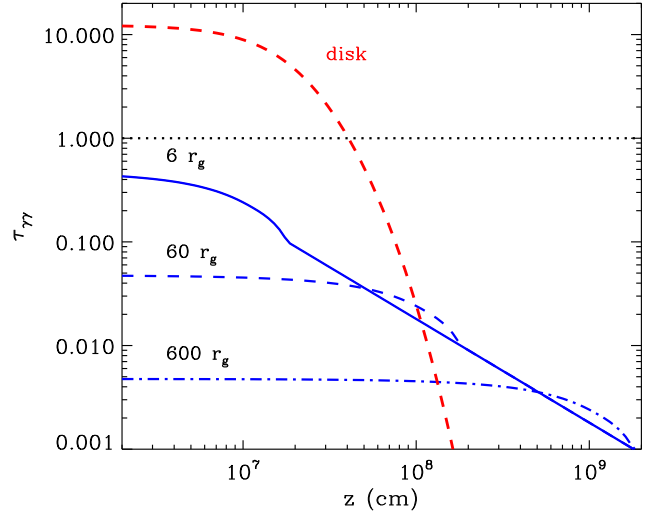}} 
  \caption{Optical depth at infinity for 1~GeV photons interacting with non-thermal hard X-rays (10-100~keV). Gamma rays are injected on the axis of the accretion disk at an altitude $z$ ($r\equiv 0$) and with an inclination $\psi=30\degr$. The corona X-ray density is homogeneous and isotropic in a spherical region of radius $R_{\rm{c}}=6~r_{\rm{g}}$ (solid line), $60~r_{\rm{g}}$ (dashed line) and $600~r_{\rm{g}}$ (dotted-dashed line) around the compact object, here a 20~\msol\ black hole ($6~r_{\rm{g}}=1.8\times 10^7~$cm). Outside, the X-ray flux is radial. The total power injected in hard X-rays is $L_{\rm{c}}=5\times 10^{36}~$erg~s$^{-1}$. The gamma-ray optical depth from the disk only is shown with the red dashed line for comparison.}
\label{abs_corona}
\end{figure}

Fig.~\ref{abs_corona} shows the contribution to the gamma-ray optical depth at 1~GeV integrated along the line of sight as a function of the location of the gamma-ray source above the accretion disk (for $r=0$). Inside the corona, $\tau_{\gamma\gamma}$ is almost constant and gamma rays are only marginally absorbed even if the corona is compact ($R_{\rm{c}} \approx R_{\rm{in}}$). Outside, the gamma-ray optical depth drops because of the decrease of the coronal photon density with distance $\propto 1/\rho^2$. Gamma-ray absorption by hard X-rays dominates over absorption by the accretion disk photons beyond a certain altitude $z$ if the corona is extended, but the optical depth is very small in this case ($\tau_{\gamma\gamma}\ll 1$) unless the luminosity in the corona is very high  $L_{\rm{c}}\gtrsim 10^{38}~$erg~s$^{-1}$. However, such high flux is not observed in hard X-rays even during major radio flares where $L_{\rm{c}}$ can be as high as $\lesssim 5\times 10^{37}$~erg~s$^{-1}$ (between 10 and 100~keV). This simple calculation shows that the non-thermal X-ray component in the soft state is not intense enough to produce an extra absorption feature in gamma rays. The results obtained in the previous section are unchanged by the presence of the corona.

\subsection{Gamma-ray emission from the corona}{\label{sect_corona_gev}}

\begin{figure}
\centering
\resizebox{\hsize}{!}{\includegraphics{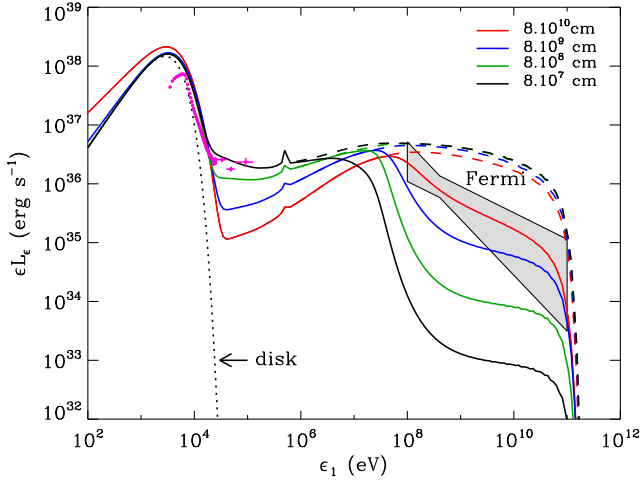}} 
  \caption{Simulated high-energy spectrum from the corona in Cygnus~X-3 with {\tt BELM}. The multicolor black body spectrum from the accretion disk (black dotted line) is comptonized by a population of thermal and non-thermal electrons heated and accelerated in the corona up to a maximum Lorentz factor $\gamma_{\rm{max}}=2\times 10^5$. The black line shows the best-fit solution (not corrected from absorption of soft X-rays by the ISM and the stellar wind) to the X-ray data (archive {\em RXTE} data, pink crosses from \citealt{2008MNRAS.388.1001S}) in the ultra-soft state given in \citet{2009MNRAS.392..251H}. This solution corresponds to a corona of radius $R_{\rm{c}}=8\times 10^{7}~$cm and a total unabsorbed bolometric luminosity $L_{\rm{bol}}=3.6\times 10^{38}$ (at 9~kpc). This figure shows also the spectra for $R_{\rm{c}}=8\times 10^{8}$ (green), $8\times 10^{9}$ (blue) and $8\times 10^{10}$~cm (red), assuming the same bolometric luminosity. The dashed lines are the intrinsic spectra (without pair production) and the solid lines are the absorbed spectra reaching the observer. The {\em Fermi} power-law taken from \citet{2009Sci...326.1512F} is overplotted for comparison.}
\label{sp_corona}
\end{figure}

In this section, we aim to model more precisely the emission from the corona in \cyg, and see whether it could explain the observed gamma-ray emission. In \citet{2009MNRAS.392..251H} the X-ray spectra of the ultra-soft state of Cygnus X-3 were fitted with the thermal/nonthermal Comptonization model {\tt EQPAIR} \citep{1999ASPC..161..375C}. In this model the X-ray source is modeled as a spherical Comptonizing cloud in which soft photons are uniformly and isotropically injected with a multicolor accretion disk spectrum. The electron distribution of the Comptonizing electrons is hybrid: at low energy the distribution of electrons is Maxwelllian with a non-thermal quasi-powerlaw extension at high energies. The Compton cloud is powered through two channels: 1) direct heating of the thermal electrons, 2) injection of high-energy electrons with a power-law distribution. The steady state particle and photon distributions in the cloud are then computed according to energy and electron/positron pair balance, as a function of the power input in the corona, as well as other parameters, such as the temperature of the soft seed photons, size of the emitting region or Thomson optical depth $\tau$ of the plasma. The calculation is performed in the one-zone approximation and radiative transfer is dealt with using an escape probability formalism.

From this analysis of the X-ray spectra of Cygnus X-3, \citet{2009MNRAS.392..251H} infer that in the ultra-soft state, about half of the coronal power is provided to the electrons in the form of non-thermal electrons injected with a slope $\Gamma_{\rm{inj}}\simeq 2.1$. In principle, these non-thermal electrons could also be responsible for the gamma-ray emission detected by {\em Fermi}. This would require that the coronal electrons are accelerated up to a $\gamma_{\rm{max}}>10^4$. In the {\tt EQPAIR} model the grids used to compute the electron and photon energy distributions are fixed and cannot be extended to an energy high enough to simulate the {\em Fermi} emission. In \citet{2009MNRAS.392..251H} the maximum Lorentz factor of the injected electrons is $\gamma_{\rm{max}}=10^3$, clearly too low to produce the gamma-ray emission. However the exact value of $ \gamma_{\rm{max}}$ is not constrained by the X-ray data and could be much larger.

In the following, we use the {\tt BELM} code \citep{2008A&A...491..617B} to estimate the gamma-ray flux predicted by this model. The {\tt BELM} code is very similar to {\tt EQPAIR} ({\em i.e.} based on the resolution of the kinetic equations in the one-zone, isotropic approximation) but offers more flexibility regarding the maximum energy of the electrons. {\tt BELM} can also take into account synchrotron emission and self-absorption (in addition to Compton, electron-proton Bremsstrahlung emission, pair production/annihilation and the internal electromagnetic cascade of pairs). We use exactly the same parameters found by \citet{2009MNRAS.392..251H} in their best fit model of the ultra-soft state except that the maximum Lorentz factor is now fixed at $\gamma_{\rm{max}}=2\times 10^5$ in order to produce gamma-ray emission up to 100~GeV. The total power injected as non-thermal electrons is set to $L_{\rm{nth}}=3\times 10^{37}$~erg~$\rm{s}^{-1}$.

We find that high-energy gamma rays are mostly produced by inverse Compton scattering. However, gamma rays suffer from a strong absorption {\em via} pair production with non-comptonized disk photons (see Fig.~\ref{sp_corona}). The gamma-ray flux cannot be explained. The radiation from the cascade of pairs in the corona is not sufficient to increase the gamma-ray flux to an observable level. Pair production can be reduced if the corona is more extended. The best fit solution to the X-ray data in the ultra-soft state corresponds to a corona of radius $R_{\rm{c}}=8\times 10^{7}$~cm~$\approx 30~r_{\rm{g}}$ (for a 20~\msol\ black hole). By keeping the total bolometric luminosity constant ($L_{\rm{bol}}=3.6\times 10^{38}$~erg~s$^{-1}$), the level of emission and the spectrum in the {\em Fermi} energy band can be reproduced if the radius of the corona $R_{\rm{c}}\gtrsim 10^{10}~\rm{cm}$ (Fig.~\ref{sp_corona}).

These results are consistent with the detailed calculation of gamma-ray absorption above the disk ({\em i.e.} inhomogeneous in the full isotropization limit, see Sect.~\ref{sect_wind}). Some spectral changes appear in X-rays because of the reduction in the compactness of the corona. Fig.~\ref{sp_corona} shows it will be difficult to reconcile the flat X-ray spectrum at 100 keV seen in very soft states (produced by electrons when the compactness is high) with significant gamma-ray emission (which requires low compactness to avoid gamma-ray attenuation by pair production). Note also that the spectrum could be changed due to anisotropic effects and inhomogeneities not taken into account in this model. It is difficult to imagine a very extended corona since most of the gravitational potential energy (90\%) remains in the region within $30~r_{\rm{g}}$ around the compact object.

The simulations in Fig.~\ref{sp_corona} assumes no magnetic field. For magnetic energy density at equipartition with the radiation in the corona, the magnetic field is about $B_{\rm{eq}}\sim 5\times 10^6~$G (for $R_{\rm{c}}=8\times 10^7~$cm). Synchrotron radiation becomes the dominant processes at high energies, but the shape of the escaping high-energy spectrum changes only slightly. The energy of the absorption cut-off remains magnetic field independent, hence the constraints on the size of the corona are unchanged. In addition, the high-energy cut-off of the intrinsic synchrotron spectrum cannot exceed $m_{\rm{e}}c^2/\alpha_{\rm F}\approx 70~$MeV (where $\alpha_{\rm F}$ is the fine structure constant, see {\em e.g.} \citealt{1983MNRAS.205..593G}) due to synchrotron losses, hence too low to account for the high-energy emission. This limit is calculated assuming no relativistic Doppler boosting and that the cooling timescale {\em via} synchrotron emission $t_{\rm{syn}}$ equals the acceleration timescale of particles in the corona which cannot be shorter than the Larmor timescale $t_{\rm{L}}$, assumption usually made {\em e.g.} in diffuse shock acceleration. This balance between acceleration and cooling implies also that it is very difficult to accelerate particles at high energies ({\em e.g.} \citealt{1984ARA&A..22..425H,2002PhRvD..66b3005A}). For the equipartition magnetic field strength, the condition $t_{\rm{syn}}\geq t_{\rm{acc}}$ gives $\gamma_{\rm{max}}\leq (9 m^2_{\rm{e}}c^4/4e^3 B_{\rm{eq}})^{1/2}= 5\times 10^4$, allowing gamma-ray emission to a few tens of GeV only. To accelerate particles to $\gamma_{\rm{max}}=2\times 10^5$, the magnetic field should not exceed $\sim 3\times 10^5~$G. These estimates are based on the optimistic assumption that the acceleration efficiency defined as $\eta=\dot{E_{\rm s}}/eBc$ equals 1 (where $\dot{E_{\rm s}}$ is the synchrotron energy losses and $e$ is the charge of the electron). The constraint on the magnetic field strength in the corona could be even more stringent than what is given here.  The magnetic field strength at equipartition is also sufficiently high to quench the development of pair cascade.

We conclude that the gamma-ray emission measured by {\em Fermi} has probably not a coronal origin.

\section{Conclusion}{\label{sect_ccl}}

High-energy gamma-ray emission in Cygnus~X-3 seems to be related to the soft X-ray state, {\em i.e.} when the X-ray spectrum is dominated by a bright accretion disk component. Gamma rays should be significantly absorbed by soft X-rays emitted by the inner (hottest) parts of the accretion disk, unless the gamma-ray source lies far enough from the disk. The apparent lack of absorption feature in the spectrum measured by {\em Fermi} indicates that high-energy gamma rays should be located at least $\sim 10^8$-$10^{10}~$cm away from the accreting star, depending on the inclination of the system. These conclusions are not affected by absorption of X-rays in the stellar wind, since this process becomes significant at larger distances $\gtrsim 10^{10}~$cm. Thomson scattering with free electrons in the wind could redistribute and isotropize the X-ray emission above the disk and affect the gamma-ray optical depth map above the accretion disk. In the limit where X-rays from the disk are fully isotropized, the gamma-ray photosphere is quasi-spherical and is located at about $10^{10}$~cm from the compact object. A more detailed treatment of this effect would require for instance Monte Carlo techniques to follow the multiple elastic scattering of X-rays in the wind.

In addition to the bright black body emission from the accretion disk, the X-ray spectrum contains a non-thermal tail in hard X-rays probably related to a corona above the disk. We modeled the corona as a spherical, isotropic and homogeneous cloud around the compact object. We showed that gamma-ray absorption with non-thermal hard X-rays from the corona is too small compared with the contribution from the disk. In this article, we present also a more precise modeling of the non-thermal radiation from the corona in Cygnus~X-3 with {\tt BELM}. This code enables to extend the energy of the electrons in the high-energy domain. Using the best fit solution to X-ray data of \citet{2009MNRAS.392..251H}, we found that gamma rays produced in the corona suffer from strong absorption by non-comptonized photons from the disk before they escape to the observer. The gamma-ray flux seen by {\em Fermi} could be reproduced if the corona is unrealistically extended ($R_{\rm{c}}\gtrsim 10^{10}~\rm{cm}$). This study does not favor a coronal origin of the observed gamma-ray emission.

At distances comparable to the orbital separation, the Wolf-Rayet star photon density dominates over the accretion disk X-ray density. The gamma-ray emission and the modulation are likely to be produced by energetic electron-positron pairs possibly located in the jet, upscattering stellar radiation {\em via} inverse Compton scattering. In this case, the \emph{Fermi} spectrum is well fitted if pairs are injected with a power-law energy distribution of index 4.4. The power-law cannot extend below $\gamma_{\rm{min}}\gtrsim 50$ (assuming no Doppler boosting) or the observed hard X-ray flux below 100~keV would be overestimated. Alternatively, the inverse Compton emission from the jet could also be responsible for the hard X-ray emission (instead of the corona). This would require a spectral break in the electron energy distribution below $\gamma_{\rm{brk}}\lesssim 10^3$ with a harder power-law of index close to 3 with $\gamma_{\rm min}\lesssim 10$ (in order to obtain a flat spectrum above 10~keV). The emerging picture is that high-energy gamma rays should be emitted far from the accretion disk, probably in the relativistic jet \citep{2010MNRAS.404L..55D}.

\begin{acknowledgements}
BC thanks Dmitri Uzdensky and Mitch~C. Begelman for the interesting discussions regarding this study. The authors are grateful to the anonymous referee for his helpful comments on the manuscript. This research was supported by the {\em European Community} via contract ERC-StG-200911, in part by the Polish MNiSW grant 362/1/N-INTEGRAL/2008/09/0 and by the GdR~PCHE.\end{acknowledgements}

\appendix

\section{Calculation of the gamma-ray optical depth in the radiation field of a standard accretion disk}\label{app_a}

\begin{figure*}
\centering
\resizebox{17cm}{!}
{\includegraphics*{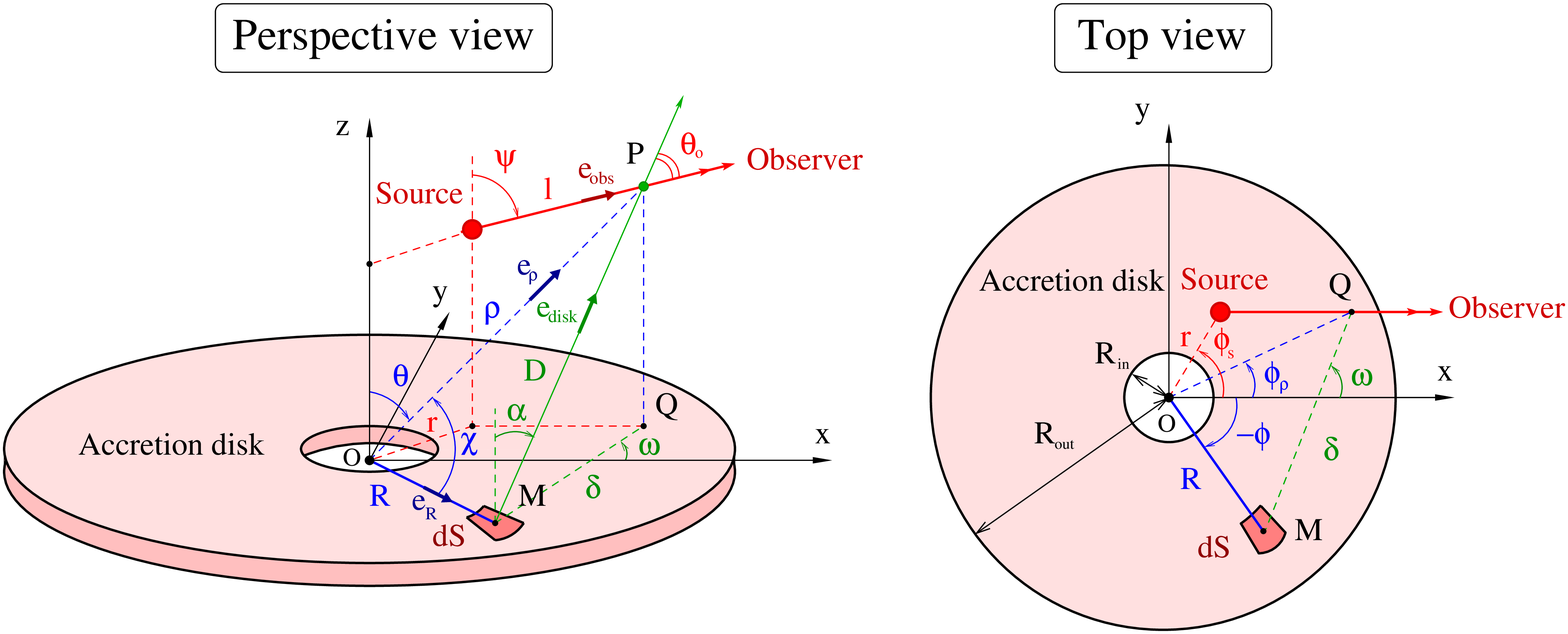}}
  \caption{Diagram of a flat, geometrically thin accretion disk formed around the compact object in the center (point O). A source of energetic gamma rays lies above the disk seen at an angle $\psi$ by a distant observer. Gamma rays and photons from the unit surface of the disk $dS$ (centered around point M) interact at the point P with an angle $\theta_0$. Point Q is the projection of point P in the disk plane. All the other geometrical quantities referred in the text for the calculation of the gamma-ray optical depth are also shown here.}
\label{fig_disk}
\end{figure*}

We aim to give here the detailed calculation of the optical depth $\tau_{\gamma\gamma}$ for a gamma-ray photon of energy $\epsilon_1$ propagating above an accretion disk. The disk is optically thick, geometrically thin and flat with an inner radius $R_{\rm{in}}$ and external radius $R_{\rm{out}}$ as depicted in Fig.~\ref{fig_disk}. The compact object lies at the center of the disk at the point O. We consider a point-like source of high-energy gamma rays located at a radial distance $r$, at an altitude $z$ above the accretion disk and with an angle $\phi_{\rm{s}}$ with the $x$-axis in the plane of the disk. A distant observer sees the disk with an angle $\psi$, this direction is set in the ($x$O$z$) plane.

\subsection{Total gamma-ray optical depth}

The total gamma-ray optical depth $\tau_{\gamma\gamma}$ integrated over the high-energy photon path length $l$ and the target photon density from the disk $dn/d\epsilon d\Omega$ (in ph~cm$^{-3}$~erg$^{-1}$~sr$^{-1}$) yields
\begin{equation}
\tau_{\gamma\gamma}=\int_{0}^{+\infty}\int_{0}^{\Omega_{\rm{disk}}}\int_{\epsilon_{\rm{min}}}^{+\infty}\frac{dn}{d\epsilon d\Omega}\left(1-\cos\theta_0\right)\sigma_{\gamma\gamma}d\epsilon d\Omega dl,
\label{tau_gg}
\end{equation}
where $\epsilon$ is the disk photon energy, $\theta_0$ the angle between the two photons, $\sigma_{\gamma\gamma}$ the pair production cross section (see {\em e.g.} \citealt{1967PhRv..155.1404G}) and $\Omega_{\rm{disk}}$ the total solid angle covered by the accretion disk seen from the gamma ray. Eq.~(\ref{tau_gg}) must also satisfy the kinematic condition
\begin{equation}
\epsilon\geq\epsilon_{\rm{min}}=\frac{2 m_{\rm{e}}^2 c^4}{\epsilon_1\left(1-\cos\theta_0\right)}
\label{threshold}
\end{equation}
where $m_{\rm{e}}$ is the rest mass of the electron (or positron), otherwise no pairs are produced {\em i.e.} $\tau_{\gamma\gamma}=0$.

\subsection{Geometrical aspects}

In order to derive explicitly the gamma-ray optical depth in Eq.~(\ref{tau_gg}), a few geometrical relations should be established. First, we define the following unit vectors in cartesian coordinates as (see Fig.~\ref{fig_disk})
\begin{equation}
\mathbf{e_{\rm{obs}}}=\left(\sin\psi,0,\cos\psi\right)
\end{equation}
along the direction joining the source and the observer,
\begin{equation}
\mathbf{e_{\rm{R}}}=\left(\cos\phi,\sin\phi,0\right)
\end{equation}
along the $\mathbf{OM}$ direction,
\begin{equation}
\mathbf{e_{\rho}}=\left(\sin\theta\cos\phi_{\rho},\sin\theta\sin\phi_{\rho},\cos\theta\right)
\end{equation}
along the $\mathbf{OP}$ direction and
\begin{equation}
\mathbf{e_{\rm{disk}}}=\left(\sin\alpha\cos\omega,\sin\alpha\sin\omega,\cos\alpha\right)
\label{e_disk}
\end{equation}
in the $\mathbf{MP}$ direction. Elementary trigonometry gives
\begin{equation}
\cos\theta=\frac{z+l\cos\psi}{\rho},\hspace{0.5cm}
\sin\theta=\frac{\left[y^2+(x+l\sin\psi)^2\right]^{1/2}}{\rho},
\label{sin_cos_theta}
\end{equation}
where $x=r\cos\phi_{\rm{s}}$ and $y=r\sin\phi_{\rm{s}}$ are the coordinates of the source in the plane of the disk. Combining the expressions given in Eq.~(\ref{sin_cos_theta}), we obtain
\begin{equation}
\rho^2=x^2+y^2+z^2+l^2+2l\left(x\sin\psi+z\cos\psi\right).
\end{equation}
The distance $D\equiv MP$ between the element surface of the disk $dS$ and the point where both photons collide is
\begin{equation}
D^2=R^2+\rho^2-2R\rho\cos\chi,
\end{equation}
with
\begin{equation}
\cos\chi=\mathbf{e_{\rho}} \cdot \mathbf{e_{\rm{R}}}=\sin\theta\cos\phi_{\rho}\cos\phi+\sin\theta\sin\phi_{\rho}\sin\phi,
\end{equation}
where
\begin{equation}
\cos\phi_{\rho}=\frac{x+l\sin\psi}{\rho\sin\theta},\hspace{0.5cm}
\sin\phi_{\rho}=\frac{y}{\rho\sin\theta}.
\label{sin_cos_phi_rho}
\end{equation}
The angle $\alpha$ between the direction $\mathbf{MP}$ and the axis of the disk $z$ introduced in Eq.~(\ref{e_disk}) is defined such as
\begin{equation}
\cos\alpha=\frac{\rho\cos\theta}{D},\hspace{0.5cm}
\sin\alpha=\frac{\delta}{D}=\frac{\left(D^2-\rho^2\cos^2\theta\right)^{1/2}}{D}.
\end{equation}
The cosine of the scattering angle $\theta_0$ between the high-energy gamma ray and the photons from the element surface of the disk is
\begin{equation}
\cos\theta_0=\mathbf{e_{\rm{obs}}}\cdot\mathbf{e_{\rm{disk}}}=\sin\psi\sin\alpha\cos\omega+\cos\psi\cos\alpha,
\end{equation}
where
\begin{equation}
\cos\omega=\frac{x+l\sin\psi-R\cos\phi}{\left(D^2-\rho^2\cos^2\theta\right)^{1/2}}.
\end{equation}
The element solid angle $d\Omega$ covered by the surface $dS=RdRd\phi$, as seen by a gamma-ray photon along its way to the observer is (see Fig.~\ref{fig_disk})
\begin{equation}
d\Omega=\frac{\mathbf{e_{\rm{disk}}}\cdot\mathbf{e_{\rm{z}}}~dS}{D^2}=\frac{\rho\cos\theta}{D^3}RdRd\phi,
\label{dOmega}
\end{equation}
$\mathbf{e_{\rm{z}}}$ is the unit vector along the $z$-axis.

\subsection{Soft photon density of the disk}

In the standard model, the accretion disk is formed by concentric annuli in thermal equilibrium emitting a black body spectrum (see {\em e.g.} \citealt{1981ARA&A..19..137P}). Hence, a surface element of the disk $dS$ generates the density
\begin{equation}
\frac{dn}{d\epsilon d\Omega}=\frac{2}{h^3 c^3}\frac{\epsilon^2}{\exp\left[\epsilon/k T(R)\right]-1},
\end{equation}
with $h$, the Planck constant, $k$ the Boltzmann constant and $T(R)$ is the effective temperature of the disk at the distance $R$ from the center. Using the temperature profile given in Eq.~(\ref{temp_profile}), the spectrum integrated over the disk is
\begin{equation}
\frac{dn}{d\epsilon}=\int_{0}^{\Omega_{\rm{disk}}}\frac{dn}{d\epsilon d\Omega} d\Omega,
\end{equation}
where $d\Omega$ is given in Eq.~(\ref{dOmega}). In the case of an observer located at a distance $L\gg R$ so that $D\approx\rho\equiv L$ and $\theta\approx\psi$, $d\Omega$ simplifies into $d\Omega\approx \cos\psi RdRd\phi/L^2$. The isotropic apparent flux of the disk seen by the observer $\epsilon L_{\epsilon}$ (in erg~s$^{-1}$) is then
\begin{equation}
\epsilon L_{\epsilon}=4\pi c L^2 \epsilon^2 \frac{dn}{d\epsilon}=8\pi^2 c \cos\psi~ \epsilon^2\int_{R_{\rm{in}}}^{R_{\rm{out}}}\frac{dn}{d\epsilon d\Omega}RdR.
\label{fnu}
\end{equation}
In the limit $R_{\rm{in}}\ll R\ll R_{\rm{out}}$, the flux can be expressed as
\begin{equation}
\epsilon L_{\epsilon}=\frac{2^{10/3} 45^{2/3} \zeta\left(\frac{8}{3}\right)\Gamma\left(\frac{8}{3}\right)G^{2/3}}{3\pi^2 h c^{2/3}}\cos\psi~\dot{M}_{\rm{acc}}^{2/3} M^{2/3}_{\rm{co}} \epsilon^{4/3},
\end{equation}
a power-law of index $4/3$ in energy. $G$ is the gravitational constant. $\zeta$ and $\Gamma$ are respectively the Riemann and the gamma functions, with $\zeta\left(8/3\right)\approx 1.2842$ and $\Gamma\left(8/3\right)\approx 1.5046$.

\section{Calculation of the gamma-ray optical depth with non-thermal hard X-rays from the corona}\label{app_b}

\subsection{Photon density of the corona}

Hard X-rays are assumed to originate from the corona, a spherical region of radius $R_{\rm{c}}$ (see Fig.~\ref{fig_corona}). The density of non-thermal X-rays follows
\begin{equation}
\frac{dn}{d\epsilon}=K_{\rm{c}} \epsilon^{-p},
\end{equation}
where $K_{\rm{c}}$ is a normalization constant and $p$ the spectral index. 
For simplicity, we assume that X-rays are emitted radially outside the corona (if $\rho> R_{\rm{c}}$). If $L_{\rm{c}}$ is the total luminosity of the corona, we have
\begin{equation}
L_{\rm{c}}=4\pi \rho^2 c \int_{\epsilon_-}^{\epsilon_+}\epsilon\frac{dn}{d\epsilon}d\epsilon.
\end{equation}
Hence the density of non-thermal X-rays outside the corona is
\begin{equation}
\frac{dn_{\rm{ext}}}{d\epsilon}\left(\rho\right)=K_0\left(\frac{R_{\rm{c}}}{\rho}\right)^2\epsilon^{-p},
\end{equation}
where (if $p\neq 2$)
\begin{equation}
K_0=\frac{\left(2-p\right)L_{\rm{c}}}{4\pi c \left(\epsilon_+^{-p+2}-\epsilon_-^{-p+2}\right)R^2_{\rm{c}}}.
\end{equation}
If X-ray photons are homogeneous and isotropic inside the corona ($\rho<R_c$), then
\begin{equation}
\frac{dn_{\rm{int}}}{d\epsilon}=\frac{dn_{\rm{ext}}}{d\epsilon}\left(\rho=R_{\rm{c}}\right)=K_0 \epsilon^{-p}.
\end{equation}

\subsection{Optical depth}

\begin{figure}
\centering
\resizebox{\hsize}{!}{\includegraphics{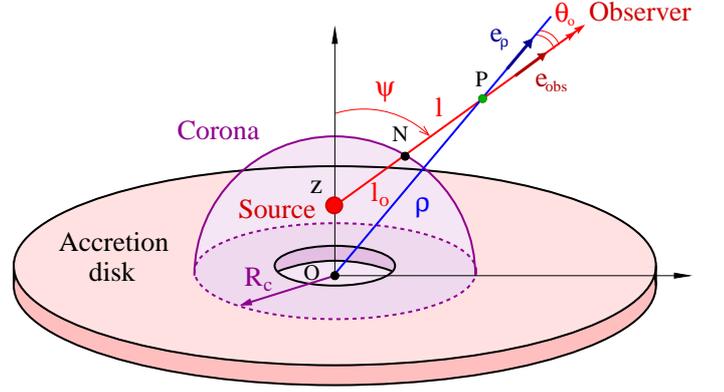}} 
  \caption{The corona is approximated as an optically thin spherical region of radius $R_{\rm{c}}$ (violet region) centered at the compact object (point O), and emit hard X-rays. Inside the corona ($\rho< R_{\rm{c}}$), the density of hard X-rays is uniform and isotropic. Outside ($\rho> R_{\rm{c}}$), photons propagate outward radially. Gamma rays are injected on the disk axis with an inclination $\psi$ towards the observer.}
\label{fig_corona}
\end{figure}

Gamma rays are injected on the axis of the accretion disk at an altitude $z$. If $z< R_{\rm{c}}$, gamma rays first cross the corona over a distance $l_0$ given by
\begin{equation}
l_0=\left[R_{\rm{c}}^2-z^2\left(1-\cos^2\psi\right)\right]^{1/2}-z\cos\psi,
\label{l0}
\end{equation}
and escape at the point N (see Fig.~\ref{fig_corona}). We average over an isotropic distribution of pitch angle between both photons $\theta_0$. The integral over the length path in the full expression of the total optical depth (Eq.~\ref{tau_gg}) is trivial since there is an invariance in $l$ for homogeneous target photon density. The gamma-ray opacity inside the corona is
\begin{equation}
\tau_{\rm{int}}\left(\epsilon_1,z\right)=\frac{l_0}{2}\int_{\epsilon_-}^{\epsilon_+}\int_{0}^{\pi}\frac{dn_{\rm{int}}}{d\epsilon}\left(1-\cos\theta_0\right)\sigma_{\gamma\gamma}\sin\theta_{0}d\theta_0 d\epsilon.
\end{equation}
Outside the corona, X-ray photons are anisotropic and not homogeneous. The angle of interaction is related to the length path $l$ {\em via}
\begin{eqnarray}
\cos\theta_0 &=& \mathbf{e_{\rm{obs}}}\cdot\mathbf{e_{\rho}} \nonumber \\
&=& \left(\sin\psi,0,\cos\psi\right)\cdot\left(\frac{(l_0+l)\sin\psi}{\rho},0,\frac{z+(l_0+l)\cos\psi}{\rho}\right) \nonumber \\
&=& \frac{l_0+l+z\cos\psi}{\rho},
\end{eqnarray}
where $l_0$ is given in Eq.~(\ref{l0}), $l$ is the distance NP, and $\rho$ is the distance to the interaction point from the center OP such that
\begin{equation}
\rho^2=(l_0+l)^2+z^2+2z\left(l_0+l\right)\cos\psi.
\end{equation}
The gamma-ray optical depth is then
\begin{equation}
\tau_{\rm{ext}}\left(\epsilon_1,z\right)=\int_{l_0}^{+\infty}\int_{\epsilon_-}^{\epsilon_+}\frac{dn_{\rm{ext}}}{d\epsilon}\left(\rho\right)\left(1-\cos\theta_0\right)\sigma_{\gamma\gamma}d\epsilon dl.
\end{equation}
The total optical depth is $\tau_{\gamma\gamma}=\tau_{\rm{int}}+\tau_{\rm{ext}}$.

\bibliographystyle{aa}
\bibliography{abscygx3}

\begin{thebibliography}{55}
\expandafter\ifx\csname natexlab\endcsname\relax\def\natexlab#1{#1}\fi

\bibitem[{{Abdo} {et~al.}(2009){Abdo}, {Ackermann}, {Ajello}, {Axelsson},
  {Baldini}, {Ballet}, {Barbiellini}, {Bastieri}, {Baughman}, {Bechtol},
  {Bellazzini}, {Berenji}, {Blandford}, {Bloom}, {Bonamente}, {Borgland},
  {Brez}, {Brigida}, {Bruel}, {Burnett}, {Buson}, {Caliandro}, {Cameron},
  {Caraveo}, {Casandjian}, {Cecchi}, {{\c C}elik}, {Chaty}, {Cheung}, {Chiang},
  {Ciprini}, {Claus}, {Cohen-Tanugi}, {Cominsky}, {Conrad}, {Corbel}, {Corbet},
  {Dermer}, {de Palma}, {Digel}, {do Couto e Silva}, {Drell}, {Dubois},
  {Dubus}, {Dumora}, {Farnier}, {Favuzzi}, {Fegan}, {Focke}, {Fortin},
  {Frailis}, {Fusco}, {Gargano}, {Gehrels}, {Germani}, {Giavitto}, {Giebels},
  {Giglietto}, {Giordano}, {Glanzman}, {Godfrey}, {Grenier}, {Grondin},
  {Grove}, {Guillemot}, {Guiriec}, {Hanabata}, {Harding}, {Hayashida}, {Hays},
  {Hill}, {Hjalmarsdotter}, {Horan}, {Hughes}, {Jackson}, {J{\'o}hannesson},
  {Johnson}, {Johnson}, {Johnson}, {Kamae}, {Katagiri}, {Kawai}, {Kerr},
  {Kn{\"o}dlseder}, {Kocian}, {Koerding}, {Kuss}, {Lande}, {Latronico},
  {Lemoine-Goumard}, {Longo}, {Loparco}, {Lott}, {Lovellette}, {Lubrano},
  {Madejski}, {Makeev}, {Marchand}, {Marelli}, {Max-Moerbeck}, {Mazziotta},
  {McColl}, {McEnery}, {Meurer}, {Michelson}, {Migliari}, {Mitthumsiri},
  {Mizuno}, {Monte}, {Monzani}, {Morselli}, {Moskalenko}, {Murgia}, {Nolan},
  {Norris}, {Nuss}, {Ohsugi}, {Omodei}, {Ong}, {Ormes}, {Paneque}, {Parent},
  {Pelassa}, {Pepe}, {Pesce-Rollins}, {Piron}, {Pooley}, {Porter},
  {Pottschmidt}, {Rain{\`o}}, {Rando}, {Ray}, {Razzano}, {Rea}, {Readhead},
  {Reimer}, {Reimer}, {Richards}, {Rochester}, {Rodriguez}, {Rodriguez},
  {Romani}, {Ryde}, {Sadrozinski}, {Sander}, {Saz Parkinson}, {Sgr{\`o}},
  {Siskind}, {Smith}, {Smith}, {Spinelli}, {Starck}, {Stevenson}, {Strickman},
  {Suson}, {Takahashi}, {Tanaka}, {Thayer}, {Thompson}, {Tibaldo}, {Tomsick},
  {Torres}, {Tosti}, {Tramacere}, {Uchiyama}, {Usher}, {Vasileiou}, {Vilchez},
  {Vitale}, {Waite}, {Wang}, {Wilms}, {Winer}, {Wood}, {Ylinen}, \&
  {Ziegler}}]{2009Sci...326.1512F}
{Abdo}, A.~A., {Ackermann}, M., {Ajello}, M., {et~al.} 2009, Science, 326, 1512

\bibitem[{{Aharonian} {et~al.}(2002){Aharonian}, {Belyanin}, {Derishev},
  {Kocharovsky}, \& {Kocharovsky}}]{2002PhRvD..66b3005A}
{Aharonian}, F.~A., {Belyanin}, A.~A., {Derishev}, E.~V., {Kocharovsky}, V.~V.,
  \& {Kocharovsky}, V.~V. 2002, \prd, 66, 023005

\bibitem[{{Akharonian} \& {Vardanian}(1985)}]{1985Ap&SS.115...31A}
{Akharonian}, F.~A. \& {Vardanian}, V.~V. 1985, \apss, 115, 31

\bibitem[{{Aleksi{\'c}} {et~al.}(2010){Aleksi{\'c}}, {Antonelli}, {Antoranz},
  {Backes}, {Baixeras}, {Barrio}, {Bastieri}, {Becerra Gonz{\'a}lez},
  {Bednarek}, {Berdyugin}, {Berger}, {Bernardini}, {Biland}, {Blanch}, {Bock},
  {Boller}, {Bonnoli}, {Bordas}, {Borla Tridon}, {Bosch-Ramon}, {Bose},
  {Braun}, {Bretz}, {Britzger}, {Camara}, {Carmona}, {Carosi}, {Colin},
  {Contreras}, {Cortina}, {Costado}, {Covino}, {Dazzi}, {De Angelis}, {De Cea
  del Pozo}, {De Lotto}, {De Maria}, {De Sabata}, {Delgado Mendez}, {Doert},
  {Dom{\'{\i}}nguez}, {Dominis Prester}, {Dorner}, {Doro}, {Elsaesser},
  {Errando}, {Ferenc}, {Fonseca}, {Font}, {Garc{\'{\i}}a L{\'o}pez},
  {Garczarczyk}, {Gaug}, {Godinovic}, {G{\"o}ebel}, {Hadasch}, {Herrero},
  {Hildebrand}, {H{\"o}hne-M{\"o}nch}, {Hose}, {Hrupec}, {Hsu}, {Jogler},
  {Klepser}, {Kr{\"a}henb{\"u}hl}, {Kranich}, {La Barbera}, {Laille},
  {Leonardo}, {Lindfors}, {Lombardi}, {Longo}, {L{\'o}pez}, {Lorenz},
  {Majumdar}, {Maneva}, {Mankuzhiyil}, {Mannheim}, {Maraschi}, {Mariotti},
  {Mart{\'{\i}}nez}, {Mazin}, {Meucci}, {Miranda}, {Mirzoyan}, {Miyamoto},
  {Mold{\'o}n}, {Moles}, {Moralejo}, {Nieto}, {Nilsson}, {Ninkovic}, {Orito},
  {Oya}, {Paiano}, {Paoletti}, {Paredes}, {Partini}, {Pasanen}, {Pascoli},
  {Pauss}, {Pegna}, {Perez-Torres}, {Persic}, {Peruzzo}, {Prada}, {Prandini},
  {Puchades}, {Puljak}, {Reichardt}, {Rhode}, {Rib{\'o}}, {Rico}, {Rissi},
  {R{\"u}gamer}, {Saggion}, {Saito}, {Saito}, {Salvati}, {S{\'a}nchez-Conde},
  {Satalecka}, {Scalzotto}, {Scapin}, {Schultz}, {Schweizer}, {Shayduk},
  {Shore}, {Sierpowska-Bartosik}, {Sillanp{\"a}{\"a}}, {Sitarek}, {Sobczynska},
  {Spanier}, {Spiro}, {Stamerra}, {Steinke}, {Struebig}, {Suric}, {Takalo},
  {Tavecchio}, {Temnikov}, {Terzic}, {Tescaro}, {Teshima}, {Torres}, {Vankov},
  {Wagner}, {Weitzel}, {Zabalza}, {Zandanel}, {Zanin}, \& {The MAGIC
  Collaboration}}]{2010ApJ...721..843A}
{Aleksi{\'c}}, J., {Antonelli}, L.~A., {Antoranz}, P., {et~al.} 2010, \apj,
  721, 843

\bibitem[{{Apparao}(1984)}]{1984ApJ...287..338A}
{Apparao}, K.~M.~V. 1984, \apj, 287, 338

\bibitem[{{Band} \& {Grindlay}(1985)}]{1985ApJ...298..128B}
{Band}, D.~L. \& {Grindlay}, J.~E. 1985, \apj, 298, 128

\bibitem[{{Becker} \& {Kafatos}(1995)}]{1995ApJ...453...83B}
{Becker}, P.~A. \& {Kafatos}, M. 1995, \apj, 453, 83

\bibitem[{{Bednarek}(1993)}]{1993A&A...278..307B}
{Bednarek}, W. 1993, \aap, 278, 307

\bibitem[{{Bednarek}(2010)}]{2010MNRAS.406..689B}
{Bednarek}, W. 2010, \mnras, 406, 689

\bibitem[{{Belmont} {et~al.}(2008){Belmont}, {Malzac}, \&
  {Marcowith}}]{2008A&A...491..617B}
{Belmont}, R., {Malzac}, J., \& {Marcowith}, A. 2008, \aap, 491, 617

\bibitem[{{Bosch-Ramon} {et~al.}(2008){Bosch-Ramon}, {Khangulyan}, \&
  {Aharonian}}]{2008A&A...489L..21B}
{Bosch-Ramon}, V., {Khangulyan}, D., \& {Aharonian}, F.~A. 2008, \aap, 489, L21

\bibitem[{{Carrami\~nana}(1992)}]{1992A&A...264..127C}
{Carrami\~nana}, A. 1992, \aap, 264, 127

\bibitem[{{Cawley} \& {Weekes}(1984)}]{1984A&A...133...80C}
{Cawley}, M.~F. \& {Weekes}, T.~C. 1984, \aap, 133, 80

\bibitem[{{Cerutti} {et~al.}(2010){Cerutti}, {Malzac}, {Dubus}, \&
  {Henri}}]{2010A&A...519A..81C}
{Cerutti}, B., {Malzac}, J., {Dubus}, G., \& {Henri}, G. 2010, \aap, 519, A81+

\bibitem[{{Coppi}(1999)}]{1999ASPC..161..375C}
{Coppi}, P.~S. 1999, in Astronomical Society of the Pacific Conference Series,
  Vol. 161, High Energy Processes in Accreting Black Holes, ed. {J.~Poutanen \&
  R.~Svensson}, 375--+

\bibitem[{{Crowther}(2007)}]{2007ARA&A..45..177C}
{Crowther}, P.~A. 2007, \araa, 45, 177

\bibitem[{{Dickey}(1983)}]{1983ApJ...273L..71D}
{Dickey}, J.~M. 1983, \apjl, 273, L71

\bibitem[{{Dubus} {et~al.}(2010){Dubus}, {Cerutti}, \&
  {Henri}}]{2010MNRAS.404L..55D}
{Dubus}, G., {Cerutti}, B., \& {Henri}, G. 2010, \mnras, 404, L55

\bibitem[{{Esin} {et~al.}(1997){Esin}, {McClintock}, \&
  {Narayan}}]{1997ApJ...489..865E}
{Esin}, A.~A., {McClintock}, J.~E., \& {Narayan}, R. 1997, \apj, 489, 865

\bibitem[{{Ford}(1984)}]{1984MNRAS.211..559F}
{Ford}, L.~H. 1984, \mnras, 211, 559

\bibitem[{{Gierli{\'n}ski} {et~al.}(1999){Gierli{\'n}ski}, {Zdziarski},
  {Poutanen}, {Coppi}, {Ebisawa}, \& {Johnson}}]{1999MNRAS.309..496G}
{Gierli{\'n}ski}, M., {Zdziarski}, A.~A., {Poutanen}, J., {et~al.} 1999,
  \mnras, 309, 496

\bibitem[{{Gould}(1979)}]{1979A&A....76..306G}
{Gould}, R.~J. 1979, \aap, 76, 306

\bibitem[{{Gould}(1983)}]{1983ApJ...274L..23G}
{Gould}, R.~J. 1983, \apjl, 274, L23

\bibitem[{{Gould} \& {Schr{\'e}der}(1967)}]{1967PhRv..155.1404G}
{Gould}, R.~J. \& {Schr{\'e}der}, G.~P. 1967, Physical Review, 155, 1404

\bibitem[{{Guilbert} {et~al.}(1983){Guilbert}, {Fabian}, \&
  {Rees}}]{1983MNRAS.205..593G}
{Guilbert}, P.~W., {Fabian}, A.~C., \& {Rees}, M.~J. 1983, \mnras, 205, 593

\bibitem[{{Hermsen} {et~al.}(1987){Hermsen}, {Bloemen}, {Jansen}, {Bennett},
  {Buccheri}, {Mastichiadis}, {Mayer-Hasselwander}, {Strong}, {Oezel}, \&
  {Pollock}}]{1987A&A...175..141H}
{Hermsen}, W., {Bloemen}, J.~B.~G.~M., {Jansen}, F.~A., {et~al.} 1987, \aap,
  175, 141

\bibitem[{{Hillas}(1984)}]{1984ARA&A..22..425H}
{Hillas}, A.~M. 1984, \araa, 22, 425

\bibitem[{{Hjalmarsdotter} {et~al.}(2008){Hjalmarsdotter}, {Zdziarski},
  {Larsson}, {Beckmann}, {McCollough}, {Hannikainen}, \&
  {Vilhu}}]{2008MNRAS.384..278H}
{Hjalmarsdotter}, L., {Zdziarski}, A.~A., {Larsson}, S., {et~al.} 2008, \mnras,
  384, 278

\bibitem[{{Hjalmarsdotter} {et~al.}(2009){Hjalmarsdotter}, {Zdziarski},
  {Szostek}, \& {Hannikainen}}]{2009MNRAS.392..251H}
{Hjalmarsdotter}, L., {Zdziarski}, A.~A., {Szostek}, A., \& {Hannikainen},
  D.~C. 2009, \mnras, 392, 251

\bibitem[{{Ling} {et~al.}(2009){Ling}, {Zhang}, \&
  {Tang}}]{2009ApJ...695.1111L}
{Ling}, Z., {Zhang}, S.~N., \& {Tang}, S. 2009, \apj, 695, 1111

\bibitem[{{Longair}(1994)}]{1994hea..book.....L}
{Longair}, M.~S. 1994, {High energy astrophysics. Vol.2: Stars, the galaxy and
  the interstellar medium}, ed. {Longair, M.~S.}

\bibitem[{{Mart{\'{\i}}} {et~al.}(2000){Mart{\'{\i}}}, {Paredes}, \&
  {Peracaula}}]{2000ApJ...545..939M}
{Mart{\'{\i}}}, J., {Paredes}, J.~M., \& {Peracaula}, M. 2000, \apj, 545, 939

\bibitem[{{Mioduszewski} {et~al.}(2001){Mioduszewski}, {Rupen}, {Hjellming},
  {Pooley}, \& {Waltman}}]{2001ApJ...553..766M}
{Mioduszewski}, A.~J., {Rupen}, M.~P., {Hjellming}, R.~M., {Pooley}, G.~G., \&
  {Waltman}, E.~B. 2001, \apj, 553, 766

\bibitem[{{Morrison} \& {McCammon}(1983)}]{1983ApJ...270..119M}
{Morrison}, R. \& {McCammon}, D. 1983, \apj, 270, 119

\bibitem[{{Moskalenko} {et~al.}(1993){Moskalenko}, {Karakula}, \&
  {Tkaczyk}}]{1993MNRAS.260..681M}
{Moskalenko}, I.~V., {Karakula}, S., \& {Tkaczyk}, W. 1993, \mnras, 260, 681

\bibitem[{{Paczy{\'n}ski}(1977)}]{1977ApJ...216..822P}
{Paczy{\'n}ski}, B. 1977, \apj, 216, 822

\bibitem[{{Paczy{\'n}ski} \& {Wiita}(1980)}]{1980A&A....88...23P}
{Paczy{\'n}ski}, B. \& {Wiita}, P.~J. 1980, \aap, 88, 23

\bibitem[{{Papaloizou} \& {Pringle}(1977)}]{1977MNRAS.181..441P}
{Papaloizou}, J. \& {Pringle}, J.~E. 1977, \mnras, 181, 441

\bibitem[{{Predehl} {et~al.}(2000){Predehl}, {Burwitz}, {Paerels}, \&
  {Tr{\"u}mper}}]{2000A&A...357L..25P}
{Predehl}, P., {Burwitz}, V., {Paerels}, F., \& {Tr{\"u}mper}, J. 2000, \aap,
  357, L25

\bibitem[{{Pringle}(1981)}]{1981ARA&A..19..137P}
{Pringle}, J.~E. 1981, \araa, 19, 137

\bibitem[{{Protheroe} \& {Stanev}(1987)}]{1987ApJ...322..838P}
{Protheroe}, R.~J. \& {Stanev}, T. 1987, \apj, 322, 838

\bibitem[{{Romero} {et~al.}(2010){Romero}, {Vieyro}, \&
  {Vila}}]{2010A&A...519A.109R}
{Romero}, G.~E., {Vieyro}, F.~L., \& {Vila}, G.~S. 2010, \aap, 519, A109+

\bibitem[{{Shakura} \& {Sunyaev}(1973)}]{1973A&A....24..337S}
{Shakura}, N.~I. \& {Sunyaev}, R.~A. 1973, \aap, 24, 337

\bibitem[{{Sitarek} \& {Bednarek}(2008)}]{2008MNRAS.391..624S}
{Sitarek}, J. \& {Bednarek}, W. 2008, \mnras, 391, 624

\bibitem[{{Sitarek} \& {Bednarek}(2010)}]{2010MNRAS.401.1983S}
{Sitarek}, J. \& {Bednarek}, W. 2010, \mnras, 401, 1983

\bibitem[{{Skinner} {et~al.}(2010){Skinner}, {Zhekov}, {G{\"u}del}, {Schmutz},
  \& {Sokal}}]{2010AJ....139..825S}
{Skinner}, S.~L., {Zhekov}, S.~A., {G{\"u}del}, M., {Schmutz}, W., \& {Sokal},
  K.~R. 2010, \aj, 139, 825

\bibitem[{{Szostek} \& {Zdziarski}(2008)}]{2008MNRAS.386..593S}
{Szostek}, A. \& {Zdziarski}, A.~A. 2008, \mnras, 386, 593

\bibitem[{{Szostek} {et~al.}(2008){Szostek}, {Zdziarski}, \&
  {McCollough}}]{2008MNRAS.388.1001S}
{Szostek}, A., {Zdziarski}, A.~A., \& {McCollough}, M.~L. 2008, \mnras, 388,
  1001

\bibitem[{{Tavani} {et~al.}(2009){Tavani}, {Bulgarelli}, {Piano}, {Sabatini},
  {Striani}, {Evangelista}, {Trois}, {Pooley}, {Trushkin}, {Nizhelskij},
  {McCollough}, {Koljonen}, {Pucella}, {Giuliani}, {Chen}, {Costa},
  {Vittorini}, {Trifoglio}, {Gianotti}, {Argan}, {Barbiellini}, {Caraveo},
  {Cattaneo}, {Cocco}, {Contessi}, {D'Ammando}, {Del Monte}, {de Paris}, {Di
  Cocco}, {di Persio}, {Donnarumma}, {Feroci}, {Ferrari}, {Fuschino}, {Galli},
  {Labanti}, {Lapshov}, {Lazzarotto}, {Lipari}, {Longo}, {Mattaini},
  {Marisaldi}, {Mastropietro}, {Mauri}, {Mereghetti}, {Morelli}, {Morselli},
  {Pacciani}, {Pellizzoni}, {Perotti}, {Picozza}, {Pilia}, {Prest},
  {Rapisarda}, {Rappoldi}, {Rossi}, {Rubini}, {Scalise}, {Soffitta},
  {Vallazza}, {Vercellone}, {Zambra}, {Zanello}, {Pittori}, {Verrecchia},
  {Giommi}, {Colafrancesco}, {Santolamazza}, {Antonelli}, \&
  {Salotti}}]{2009Natur.462..620T}
{Tavani}, M., {Bulgarelli}, A., {Piano}, G., {et~al.} 2009, \nat, 462, 620

\bibitem[{{van Kerkwijk} {et~al.}(1996){van Kerkwijk}, {Geballe}, {King}, {van
  der Klis}, \& {van Paradijs}}]{1996A&A...314..521V}
{van Kerkwijk}, M.~H., {Geballe}, T.~R., {King}, D.~L., {van der Klis}, M., \&
  {van Paradijs}, J. 1996, \aap, 314, 521

\bibitem[{{Vestrand}(1983)}]{1983ApJ...271..304V}
{Vestrand}, W.~T. 1983, \apj, 271, 304

\bibitem[{{Vilhu} {et~al.}(2009){Vilhu}, {Hakala}, {Hannikainen}, {McCollough},
  \& {Koljonen}}]{2009A&A...501..679V}
{Vilhu}, O., {Hakala}, P., {Hannikainen}, D.~C., {McCollough}, M., \&
  {Koljonen}, K. 2009, \aap, 501, 679

\bibitem[{{Wu} {et~al.}(1993){Wu}, {Zhang}, \& {Li}}]{1993Ap&SS.204..141W}
{Wu}, M., {Zhang}, C., \& {Li}, T. 1993, \apss, 204, 141

\bibitem[{{Yakovlev} \& {Pethick}(2004)}]{2004ARA&A..42..169Y}
{Yakovlev}, D.~G. \& {Pethick}, C.~J. 2004, \araa, 42, 169

\bibitem[{{Zhang} \& {Cheng}(1997)}]{1997ApJ...475..534Z}
{Zhang}, L. \& {Cheng}, K.~S. 1997, \apj, 475, 534

\end{thebibliography}

\end{document}